\documentclass[dvips]{article}
\usepackage{icrctc07}

\title{Radio Detection of Neutrinos from Behind a Mountain}
\shorttitle{Radio detection of neutrinos}
\authors{O. Brusova$^{1}$, L. Anchordoqui$^{2}$, T. Huege$^{3}$, K. Martens$^{1}$.}
\shortauthors{O. Brusova et al.}
\afiliations{$^1$University of Utah, Department of Physics, 115S 1400E, Salt Lake City, UT 84112, USA\\
$^2$Department of Physics,
University of Wisconsin-Milwaukee, P.O. Box 413, Milwaukee, WI 53201, USA\\
$^3$Institut f\"ur Kernphysik, Forschungszentrum Karlsruhe, Postfach 3640, 76021 Karlsruhe, Germany}
\email{brusovao@physics.utah.edu}

\abstract{
We explore the  sensitivity of a neutrino detector employing strongly 
directional high gain radio antennae to detect the conversion of neutrinos 
above $10^{16}$ eV in a mountain or the earth crust. The directionality of the 
antennae will allow both, the low threshold and the suppression of background. 
This technology would have the advantage that it does not require a suitable 
atmosphere as optical detectors do and could therefore be deployed at any 
promising place on the planet. In particular one could choose suitable 
topographies at latitudes that are matched to promising source candidates.}
\begin{document}
\maketitle
%Begin the section.

\section{Introduction}

Identifying the sources of the highest energy Cosmic Rays (CR) is a
fundamental and unsolved problem in astroparticle physics.  If they are 
accelerated in astrophysical objects, high energy neutrino emission 
should be associated with this acceleration~\cite{Halzen:2002pg}. 
Since these neutrinos would neither interact with intergalactic or 
interstellar media nor be deflected by magnetic fields, neutrino 
astronomy carries the hope of identifying these elusive sources of 
highest energy CRs.

Various detectors have been proposed for neutrino energies above 10$^{16}$~eV, 
most of them relying on mountains to provide suitable target mass for 
$\nu_\tau$ with the ensuing $\tau$ lepton decay providing a detectable air 
shower (AS) after the $\tau$ leaves the target~\cite{Fargion:2000iz}. 
As high energy neutrinos get absorbed in the earth, geometries for such 
events are restricted to within a few degrees of horizontal~\cite{kaiICRC07}.
%At energies above 10$^{19}$~eV the $\tau$ decay length in vacuum exceeds 500~km, increasingly putting these decays out of the reach of detectors. 

Optical detectors exploiting the fluorescence emission of horizontal showers 
as well scintillator telescopes have been proposed to measure the promising 
$\nu_\tau$ signature~\cite{Cao:2004sd}. Here we explore what 
could be a simpler and cheaper technology in the field: strongly 
directional (high gain, low threshold) radio antennas. All they require is 
a rigid mechanical support structure and like all the other detectors some 
electronics. But no optical components have to be calibrated and maintained, 
and the detector could be made largely insensitive to the local climate at 
a promising detector site. 
The challenge for our detector will lie in isolating the signal. 

\section{Geosynchrotron radio signals}

AS produce 
geosynchrotron radio emission due to the deflection of secondary shower 
electrons and positrons in the earth's magnetic field~\cite{Huege:2003up}.
In this study for the first time such radio signals are studied from 
horizontal showers. 
Using the REAS2 Monte Carlo code~\cite{Huege:2006kd} we 
calculate the radio emission from $10^{15}$~eV and $10^{16}$~eV
horizontal, $\pi^{+}$ induced AS
propagating through a constant density atmosphere of 
$1.058 \times 10^{-3}$~g~cm$^{-3}$, corresponding to approximately 1500~m 
above sea level. The magnetic field was set to 
a strength of 0.4~Gauss with an inclination of $60^{\circ}$, a conservative 
estimate for Central Europe and Northern America. 
The AS were simulated as propagating from north to south; the corresponding 
geomagnetic angle is thus $60^{\circ}$. 
This is the most conservative choice, as it effectively minimizes the 
field component perpendicular to the propagation direction of the AS. 
Atmospheric curvature can be neglected 
here, as the AS need only a few kilometers to develop 
in the lower atmosphere. 
The AS simulations were carried out using CORSIKA 6.502 
~\cite{HeckKnappCapdevielle1998}, with shower-to-shower fluctuations taken out 
by simulating 25~air showers per parameter set and then selecting a shower 
with a typical longitudinal evolution profile for the radio simulations.

\begin{figure}[!htb]
\centering
\includegraphics[width=4.80cm,angle=270]{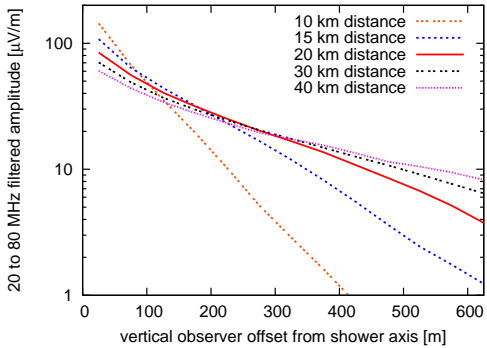}
\caption{Lateral dependence of geosynchrotron radio emission field strenghts 
of $10^{16}$~eV horizontal air showers for different observer distances from 
the $\tau$ decay point. $10^{15}$~eV values are lower by approximately a 
factor of 10.}\label{fig:distances}
\end{figure}

One important question is which observer distance is suited best for radio 
observations of these horizontal AS. 
Figure~\ref{fig:distances} illustrates how the lateral profile of the
radio signal spreads out as the distance from the starting point of
the AS increases. Despite strong relativistic beaming of
geosynchrotron radiation, the lateral slope changes significantly with the
distance of the observer from the $\tau$ decay point. An observing
distance of 20~km seems to constitute a good compromise: the signal
in the center region is still strong while its lateral spread does not 
require too dense an array of antennae. 

\begin{figure}[!htb]
\centering
\includegraphics[height=4.80cm]{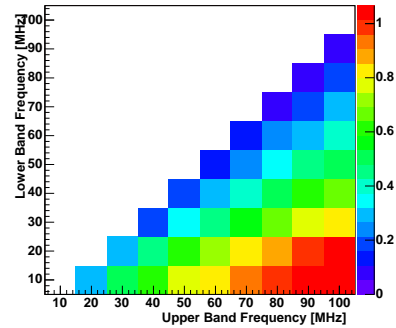}
\caption{Expected SNR as a function of observing frequency 
band for $10^{16}$~eV showers initiated at 20~km distance and 225~m vertical 
observer offset from the shower axis, calculated for an isotropic radiator. 
$10^{15}$~eV values are lower by approximately a factor of 100.}
\label{fig:freqsnr}
\end{figure}

Also important is the selection of a suitable observation frequency
band. This clearly depends on the actual noise situation at the
selected observing site. In the absence of man-made radio % frequency
interference, atmospheric and galactic noise set the limits. In
Fig.~\ref{fig:freqsnr} we show what signal-to-noise ratios (SNR,
defined as peak power of the signal divided by power of the noise in
the band of interest) can be expected for $10^{16}$~eV AS at
20~km observing distance and 225~m vertical observer offset from the
shower axis. The values are calculated for an isotropic radiator and a
combination of day-time atmospheric noise and galactic noise based on
measurements by the CCIR/ITU-R (International Telecommunication
Union). As geosynchrotron emission has a steeply falling
frequency spectrum, it is important to include low frequencies. 
As can be seen in Fig.~\ref{fig:freqsnr} an observing 
bandwidth from 20 to 80~MHz would be desirable. Below 20~MHz, atmospheric 
noise gets very strong, and above 80~MHz FM radio transmitters could pose 
problems.

\begin{figure}[!htb]
\centering
\includegraphics[width=4.80cm,angle=270]{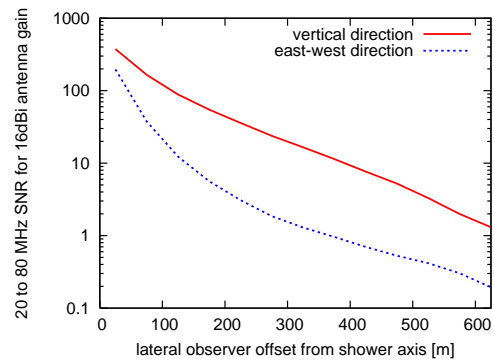}
\caption{Expected SNR as a function of observing frequency 
band for $10^{16}$~eV showers at 20~km distance, calculated for an antenna 
with 16~dBi antenna gain. $10^{15}$~eV values are lower by approximately a 
factor of 100.}
\label{fig:lateralsnr}
\end{figure}

SNR of order unity are too low for self-triggered 
measurements of geosynchroton radiation. Using directional 
antennae pointing at the target mass will significantly improve 
the SNR. For the envisioned 
broad-band measurements, logarithmic-periodic dipole antennae (LPDAs) can 
achieve antenna gains of 10~dBi. If three such antennae are phase coupled,
effective gains of up to 16~dBi can be reached. This boosts the 
SNR into a region where measurements seem feasible up to 
axis offsets of $\sim 300$~m, as illustrated in Fig.\ \ref{fig:lateralsnr}. 

For the specific relative geometry of AS and magnetic field explored 
here the lateral distribution of the radio signal is very different along 
the horizonatal and the vertical axis. 
As multiple measurements along the vertical also allow to place additional 
constraints on the zenith angle of the observed shower and as the signal falls 
off more slowly along the vertical axis, 
more than one antenna should be used along the vertical direction. 

The SNR required for self-triggered operation of an antenna array depends on 
many factors such as the multiplicity of antennae used in coincidence and the 
required detection efficiencies.
These issues would have to be addressed in a specific proposal for such a 
detector. 

\section{Neutrino event rates}

A threshold close to 10$^{16}$~eV for horizontal AS is well matched to the 
effective threshold beyond which $\tau$ leptons can escape from significant 
depth inside a rock mass: at 10$^{15}$~eV the decay length of a $\tau$ 
(in vacuum) is only $\sim$50~m. 
At 10$^{17}$~eV this decay length will have grown to 5~km, 
which allows for $\tau$ from a reasonable amount of target mass to escape 
the rock before they decay, and on the other hand constrains the 
decay volume needed behind the rock to be reasonably small. 

Using a modified ANIS~\cite{Gazizov:2004va} code we examine the
probability that a $\nu_\tau$ interacts in rock with the ensuing
$\tau$ decay initiating an AS outside of the rock. In ANIS we use the
option of CTEQ5 deduced cross-sections.  A smooth approximation for
the energy loss of the $\tau$ leptons in rock and air is included in
the calculation.  Depending on the energy of the incoming $\nu_\tau$,
an equilibrium is reached at some point inside the rock between
$\nu_\tau$ interactions producing new $\tau$ leptons and the decay of
$\tau$ leptons that were produced in neutrino interactions further
upstream. 
Our simulations show that for standard rock this
equilibrium is reached after 1~km at 10$^{15}$~eV, 2~km at
10$^{16}$~eV, and 10~km at 10$^{17}$~eV.

\begin{figure}[!htb]
\centering
\includegraphics[width=6.0cm]{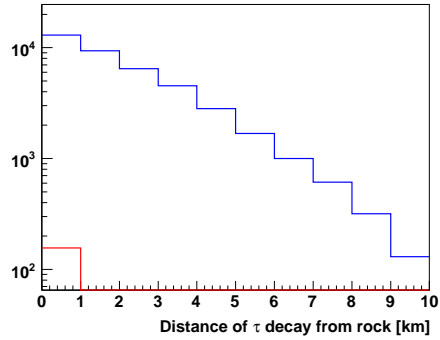}
\caption{Distribution of decay vertices that result in showers with an energy 
above 8$\times$10$^{15}$~eV. The blue histogram is for a $\nu_\tau$ energy of 
10$^{17}$~eV, the red one for 10$^{16}$~eV. 
}
\label{fig:tdec}
\end{figure}

25~M $\nu_\tau$ were injected at each of the energies mentioned above. 
From the above simulations of radio signals we infer an effective threshold of 
8$\times$10$^{15}$~eV for AS detection to estimate how many neutrinos we might 
detect. Fig.~\ref{fig:tdec} shows the distributions of decay vertices above 
that energy from the 10$^{16}$~eV and 10$^{17}$~eV simulations. 

Within the first 10~km after the mountain we see less than 200 decays
for the lower $\nu_\tau$ energy and about 40,000 for the higher one.
At higher energies efficiency will be affected by $\tau$ decaying
behind the detector ``volume''.  The detection efficiency for
10$^{17}$~eV $\nu_\tau$ is $\epsilon = 0.0016.$ A detector element of
two stations separated vertically by 200~m would allow to collect data
over a vertical range of 400~m. Working with a 200m horizontal
displacement between such detector elements one would hope to have a
decent efficiency for threefold coincidences over an area of roughly
160,000~m$^2$ for two such elements, and each vertical expansion of
the array by one additional element would add roughly 80,000~m$^2$.
The exposure of an eight antenna array accumulated over one year would
be $31,536,000~{\rm s} \times 320,000~{\rm m}^2 
  \approx  10^{17}~{\rm cm}^2\ {\rm s}$
for a pointlike source. As the LDPAs typically have an opening angle
of $65^\circ$, detectors situated 30~km from a 1.5~km high mountain
range would cover a solid angle of $5.7 \times 10^{-2}$~sr.

\begin{figure}[!htb]
\centering
\includegraphics[width=6.0cm]{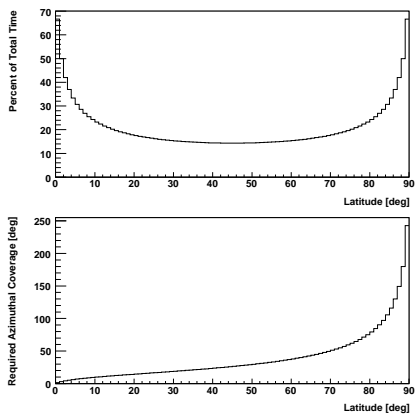}
\caption{Upper panel: Percent of total time spent in a $\pm 1.5^\circ$ band 
around the horizon if the upper or lower culmination point is arranged to be 
$1.5^\circ$ above or below the horizon. 
Lower panel: total number of degrees that have to be covered in azimuth 
in order to attain that maximal time.}
\label{fig:azlat}
\end{figure}

A source would only be seen if it is behind a target mass at the local
horizon of the detector. 
%For a given declination of a prospective
%source this singles out two specific latitudes for which the source
%will be culminating behind the horizon towards either the north or the
%south of the detector. 
%The particular magnetic field configuration
%chosen above would therefore be applicable to this case.
% Since such a latitude constraint limits the choice of geographic locations, 
% a suitable detector should be immune to the specifics of the local climate, 
% % should work 24 hours a day, 
% and be as maintenance free and rugged as possible. 
% This is a new element in the discussion of possible high energy neutrino 
% detectors and leads us to explore radio detection with strongly 
% directional, i.e. high gain antennae. 
% The amount of local radio interference will affect the detector performance. 
The time any given source spends near the horizon depends on its
declination.  
%A source at 0 degree would always be at the horizon at
%either of the poles, but in order to exploit this the detector would
%have to cover $360^\circ$ in azimuth.  From the equator on the other
%hand the north or south celestial pole would not be moving at all. 
%Apart from this peculiar point, the moving of the source behind the horizon 
%would naturally allow off-source background subtraction. 
Figure~\ref{fig:azlat} summarizes the total observation times that can
be expected if the source can be followed within $\pm 1.5^\circ$ of
the horizon and gives the azimuthal coverage required to follow the
source along that segment of its path.  If a source were to just rise
or set vertically through a $\pm 1.5^\circ$ detector aperture, it would
be visible for less than 1\% of the total time.  If the location was
chosen to accommodate the requisite azimuthal coverage shown in
Fig.~\ref{fig:azlat}, then one would expect to ``see'' the source
about 15\% of total time for a latitude of $45^\circ$.  The proposed
antennae will cover the requisite $30^\circ$ in azimuth. Putting all
this together, if we can measure $\nu_\tau$ between 10$^{16}$~eV and
10$^{17}$~eV, with the efficiencies estimated above and linearly
interpolated between the two energies, it would take an array of a
little more than 130k antennae to observe one event per year from Galactic
sources~\cite{Kistler:2006hp}.

To give an example of the sensitivity reach to the diffuse neutrino
flux we consider, $\phi_{\nu_\tau} \simeq 10^{-3}\, (E_\nu/{\rm GeV}
)^{-2.54}~{\rm GeV}^{-1}\, {\rm cm}^{-2} \,{\rm s}^{-1} \,{\rm
  sr}^{-1},$ which is expected if extragalactic cosmic rays (from
transparent sources) begin dominating the observed spectrum at
energies as low as $\sim 10^{17.6}~{\rm eV}$~\cite{Ahlers:2005sn}, as
suggested by recent HiRes data~\cite{Abbasi:2002ta}. For such a
$\nu_\tau$-flux, the expected number of events (with $E_\nu
>10^{17}$~eV) per year per eight antennae is about 0.003.

\section{Conclusions}

The energy range at which the detector works is well matched to the 
problem of $\nu_\tau$ detection through $\tau$ decay in the atmosphere. 
Cosmic ray detectors with upward of a thousand detector stations are proven,
keeping open in principle the possibility of deploying an antenna array
big enough for the isotropic fluxes. To further pursue this route, emphasis
will have to be put on developing inexpensive and reliable detector stations 
for such a detector. Radio detection carries that promise.

\section{Acknowledgements}

We thank O. Kroemer for very useful discussions.

\end{document}